\documentclass{IEEEtran}
\usepackage{cite}
\usepackage{amsmath,amssymb,amsfonts}
\usepackage{graphicx}
\usepackage{textcomp,nicefrac}
\usepackage{url}
\usepackage{booktabs}
\usepackage{hyperref}
\hypersetup{hidelinks} 
\def\BibTeX{{\rm B\kern-.05em{\sc i\kern-.025em b}\kern-.08em
T\kern-.1667em\lower.7ex\hbox{E}\kern-.125emX}}
\begin{document}
\title{A Compact Readout Electronics Based on Current Amplifier for Micromegas Detector\\ in Muon Imaging}
\author{Ting Wang, Yu Wang*, Zhihang Yao, Yulin Liu, Changqing Feng, Zhiyong Zhang and Shubin Liu
\thanks{Manuscript received 20 May 2024; revised 6 August 2024 and 27 September 2024. This work was supported by the Young Scientists Fund of the National Natural Science Foundation (Grant No. 12205297) 
and the National Science Fund for Distinguished Young Scholars (Grant No. 12025504) (Corresponding Author Yu Wang).}
\thanks{Ting Wang, Yu Wang, Zhihang Yao, Yulin Liu, Changqing Feng, Zhiyong Zhang, and Shubin Liu are with  State Key Laboratory of Particle Detection and Electronics, University of Science and Technology of China, Hefei 230026, China. (e-mail: wyu0725@ustc.edu.cn)\par
Ting Wang, Yu Wang, Yulin Liu, Changqing Feng, Zhiyong Zhang, and Shubin Liu are also with the Department of Modern Physics, University of Science and Technology of China, Hefei 230026, China.\par
Zhihang Yao and Shubin Liu are also with School of Nuclear Science and Technology, University of Science and Technology of China, Hefei 230026, China.}}

\maketitle

\begin{abstract}
Muon imaging technology is an innovative imaging technique that can be applied in volcano imaging, heavy nuclear material detection, and archaeological research. The Micromegas detector is a promising choice for muon imaging due to its high spatial resolution and large area. However, the large number of readout channels poses a challenge for electronics. In this paper, a compact front-end electronics (FEE) for reading Micromegas detectors is presented. The electronics use the commercial current-to-digital readout chip, ADAS1128, which integrates 128 current amplifiers for multi-channel charge measurement. After verifying the basic performance of the electronics, the energy resolution was obtained with a radioactive source. Furthermore, a muon imaging system prototype was set up and its spatial resolution was evaluated in a test with cosmic ray muons. The system prototype can reconstruct the boundaries of sufficiently massive objects with a size of 2 cm in a scattering imaging test.
\end{abstract}


\section{Introduction}
\label{sec:introduction}
\IEEEPARstart {M}{uon} imaging is a new type of imaging technology divided into three categories: scattering, transmission, and muon metrology. Cosmic ray muons have strong penetrating power and are universally present on Earth, which means they can be used to image the internal structure of large and shielded objects in a non-destructive and safe way. Currently, muon imaging technology has been used in the fields of volcano imaging~\cite{bib:bib1}, heavy nuclear material detection~\cite{bib:bib2}, and archaeological research~\cite{bib:bib3}. 
In recent years, a variety of detection methods based on different detectors have been applied to muon imaging, for instance, plastic scintillators~\cite{bib:bib19}, resistive plate chambers (RPCs)~\cite{bib:bib20}, multi-gap resistive plates (MRPCs)~\cite{bib:bib21}, gas electron multipliers (GEMs)~\cite{bib:bib22} and Micromegas~\cite{bib:bib14}. These detection methods have certain advantages and disadvantages in terms of detection area, spatial resolution, and scalability. 

With the continuous advancement of microelectronics and printed circuit board manufacturing processes, the area of Micro-Pattern Gaseous Detectors (MPGD) has been increasing while the readout units have become finer, particularly in the case of Micromegas detectors. It has been shown that Micromegas detectors can achieve an effective area of more than 1 m $\times$ 1 m~\cite{bib:bib4} and a spatial resolution better than 200 $\mu$m~\cite{bib:bib5}. In response to the need for multi-channel readout of large-area Micromegas detectors, several specialized Application-Specific Integrated Circuit (ASIC) chips have been developed. For example, MICROROC~\cite{bib:bib6}, a digital chip designed by the French Omega group, AGET (Asic for General Electronics for Tpc)~\cite{bib:bib7}, an analog chip developed by the French CEA Scalay Laboratory, and VMM~\cite{bib:bib8}, an analog chip designed by Brookhaven National Laboratory (USA). 

Dedicated ASICs are commonly designed with complex functions such as waveform output, time measurement, and trigger judgment to meet the readout requirements of particle physics experiments. Consequently, these designs may incur higher costs and require larger package areas to fulfill specific design requirements. In contrast, some commercial chips simplify output information to reduce costs and improve chip integration. Despite this simplification, they can still effectively read detector signals in certain applications. For example, the highly integrated commercial current readout chip ADAS1128 from Analog Devices is used to read out gas ionization chambers~\cite{bib:bib23}. This commercial current readout chip outputs charge information and has a long integration time, which makes it unsuitable for reading out Micromegas detectors at high events rate. In muon imaging experiments, the primary requirement is to measure the position of muon hits in the detector. Moreover, the rate of natural muon events is low~\cite{bib:bib24}, and the interval between the arrival of two muon events is larger than the integration time of the ADAS1128 chip. Therefore, in cosmic ray muon imaging usage scenarios, the above chip is adequate to read out Micromegas detectors.

In this paper, we describe the design of compact readout electronics based on commercially available front-end chips for highly segmented Micromegas detectors. To validate the performance of these chips in reading out Micromegas detectors for muon imaging, a small-scale prototype was constructed and tested.

\section{Micromegas Detectors}
\begin{figure}[t]
\centerline{\includegraphics[width=3.5in]{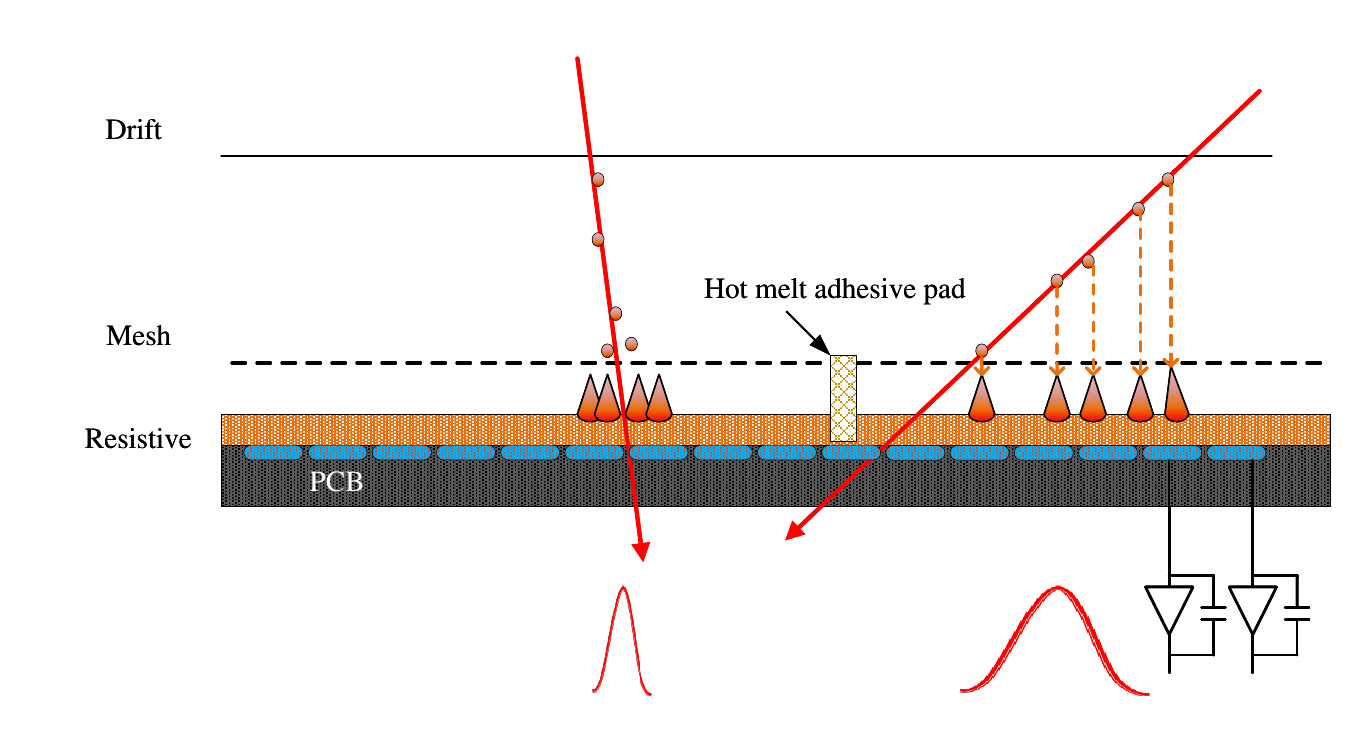}}
\caption{Diagram of Micromegas detectors.}
\label{fig1}
\end{figure}

The Micromegas detector used in this paper was designed by the author's laboratory~\cite{bib:bib9}. It primarily consists of anode readout strips, a resistive electrode, a mesh electrode, a drift electrode, and a gas chamber~\cite{bib:bib10}, as shown in \figurename~\ref{fig1}. The avalanche area is formed between the mesh electrode and the resistive electrode, and the drift area is formed between the drift electrode and the mesh electrode. The detector is read out by crossed anode strips in both X and Y directions with a pitch of 400 $\mu$m. The effective area of this Micromegas detector is 15 $\times$ 15 cm$^2$ and it has 768 channels. In our application, the working gas of the detector is a mixture of 93\% Ar and 7\% CO$_{2}$.\par
\begin{figure}[t]
\centerline{\includegraphics[width=3.5in]{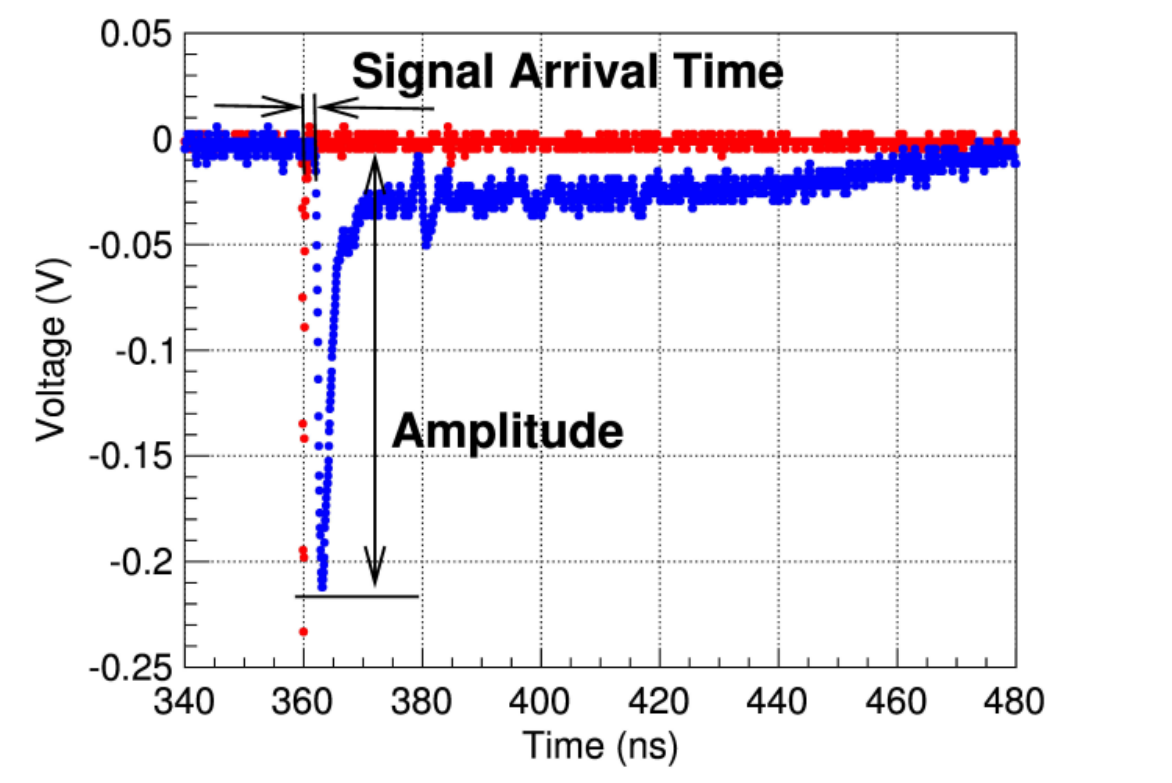}} 
\caption{Micromegas detector anode output signal. The blue line is the Micromegas detector output signal, and the red line is the timing reference of the microchannel plate MCP signal~\cite{bib:bib11}.}
\label{fig2}
\end{figure}

 When charged particles enter the detector, they produce ions and electrons in the drift region. When these primary electrons enter the avalanche region, they generate a large number of electron-ion pairs. These electrons drift rapidly toward the anode region, inducing a fast signal at the nanosecond scale. The positive ions drift towards the mesh, inducing a slow signal at hundred-nanosecond scale. Therefore, the signal obtained by the anode strips is composed of a fast electron signal and a slower ion signal superimposed~\cite{bib:bib11}, as shown in \figurename~\ref{fig2}.

For cosmic ray muon events, the pulse width of the output signal from the anode strips of the Micromegas detector is approximately 100 ns, and the output charge can typically reach a few hundred fC. To accurately measure the information of charged particle incidents on the Micromegas detector and perform readout of a large number of detector units, the electronics system needs to fulfill the requirements of low noise and high integration.

\section{Readout Electronics Design}

\subsection{Design of Readout Electronics System}
\begin{figure}[t]
\centerline{\includegraphics[width=3.5in]{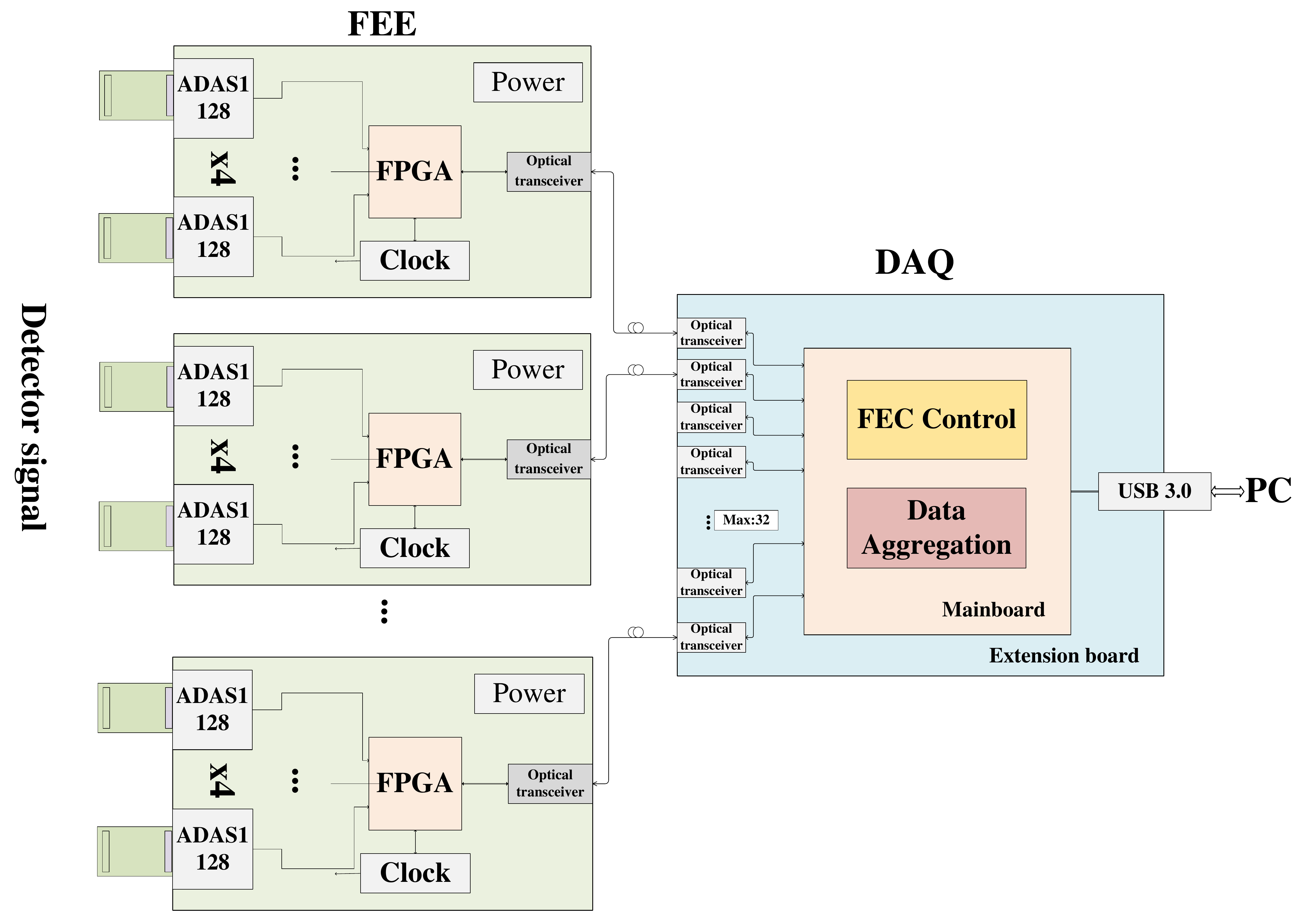}}
\caption{Architecture of the readout electronics system.}
\label{fig6}
\end{figure}

The architecture of the scalable readout electronics system used in this paper is shown in \figurename~\ref{fig6}. It comprises highly integrated front-end electronic (FEE) boards, a data acquisition (DAQ) board, and a PC. Each FEE board amplifies and digitizes detector signals, and transmits these data to the DAQ board via an optical fiber link running at 200 Mbps. The DAQ board processes and aggregates the data from multiple FEE boards and sends them to the PC via a USB 3.0 link at a rate up to 190 MB/s for final processing and storage.
The DAQ board is a generic back-end board designed in our laboratory for various applications, including muon imaging experiments~\cite{bib:bib12}. With this structure, this paper presents the FEE board built with six ADAS1128 chips and demonstrates its ability to acquire signals from Micromegas detectors.

\subsection{Introduction to the ADAS1128}
\begin{figure}[t]
\centerline{\includegraphics[width=3.5in]{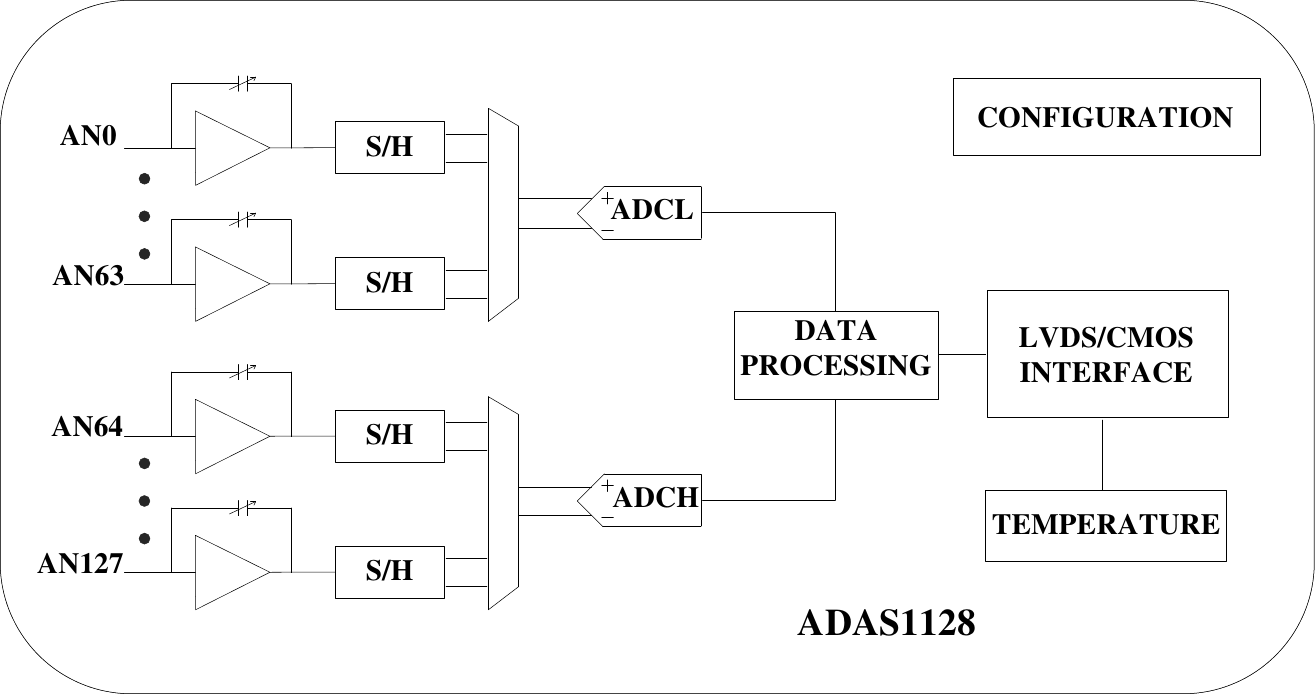}}
\caption{Diagram of ADAS1128 chip.}
\label{fig3}
\end{figure}

The FEE adopts the current readout chip ADAS1128~\cite{bib:bib13} designed by ADI. 
A simplified internal structure of this device is shown in \figurename~\ref{fig3}. The chip is composed of 128 current integration amplifiers, sample circuits, two 24-bit resolution analog-to-digital converters (ADCs), a data processing part, and two 24-bit configuration registers. 
Each input current signal is integrated by an integrating amplifier and sampled by a sample circuit. The output signal is fed to the built-in ADCs for analog-to-digital conversion.
The digital code values output from the ADCs are transmitted through the Low-Voltage Differential Signaling (LVDS) serial interface. 

The integration time can be flexibly configured from 50.7 $\mu$s to 900 $\mu$s. The dynamic range is related to the sampling rate and the gain of the integrator. As a current-to-digital readout chip, the integration time of the ADAS1128 is the same as the readout cycle time. The adjustable dynamic range is -660.17 fC to 32.23 pC when the integration time is 50.7 $\mu$s.
The dynamic range and the integration time are configurable via a 4-wire serial peripheral interface (SPI). Moreover, The chip is mounted in a 1 cm $\times$ 1 cm BGA (Ball Grid Array) package and has a power consumption of 4.5 mW per channel.\par
After power-up, the built-in automatic procedure for calibrating the gain and offset of the 128 measurement channels can be initiated via SPI. The chip then operates in continuous acquisition mode, integrating all charge signals during the specific integration time. Once a single integration cycle is completed, a new integration cycle begins. The conversion results for the 128 channels from the previous cycle (N) are output in the second subsequent integration cycle (N+2). Thus, this chip operates without dead time and charge loss. 

To read out Micromegas detectors for muon imaging, the integration time of the chip is set to 50.7 $\mu$s (the shortest integration time) to prevent recording more than one muon event within an integration time window. According to the manufacturer's datasheet, the inherent noise of the chip increases when the dynamic range is increased while the integration time remains fixed. Therefore, the dynamic range of the chip is set to -77.97 fC to 0 fC (ADC code values from 0 to 16383) to cover the signal range of the detector while meeting the low noise readout requirement.

\subsection{Design of FEE}
\begin{figure}[t]
\centerline{\includegraphics[width=3.5in]{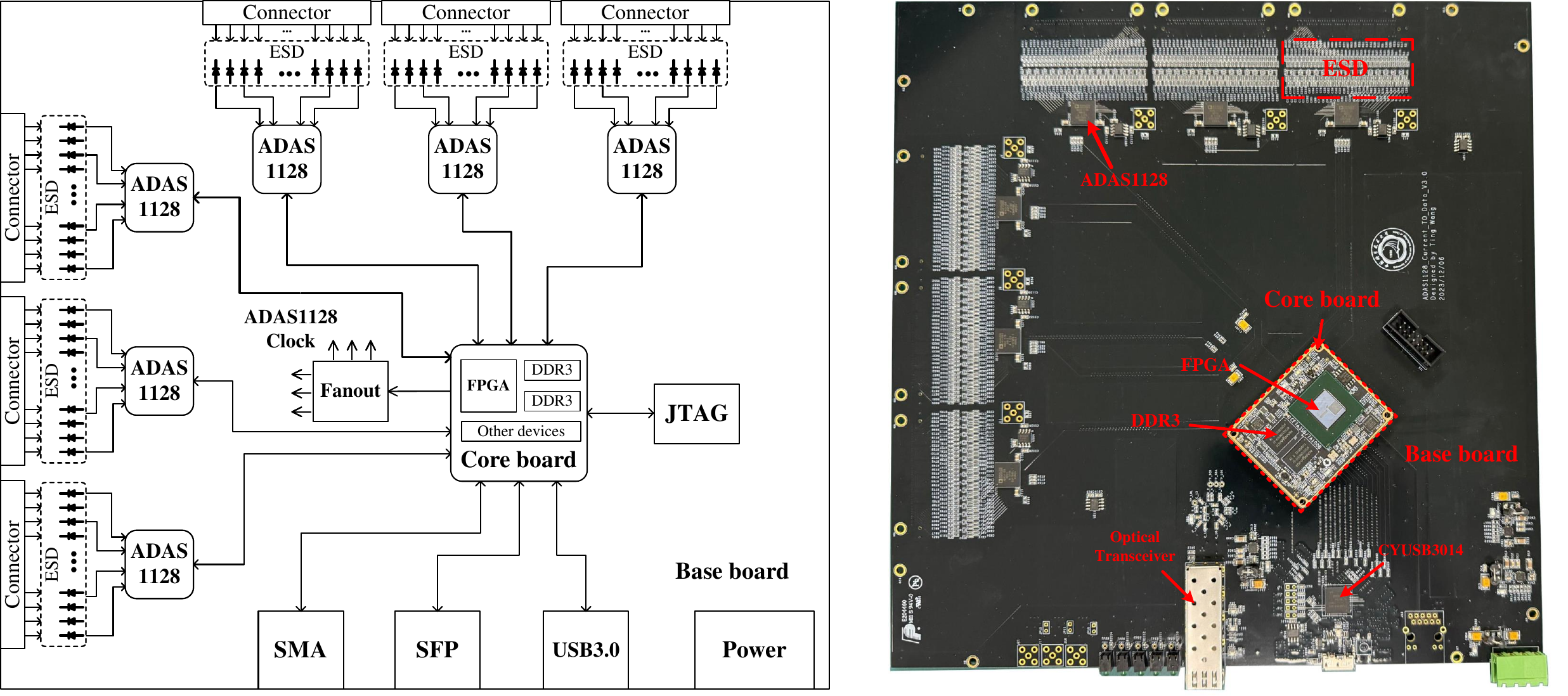}}
\caption{Schematic diagram and photograph of the FEE board.}
\label{fig4}
\end{figure}

The FEE consists of a ready-to-use module (core board) and a base board, as shown in \figurename~\ref{fig4}. The core board and the base board are interconnected via four 0.8 mm-pitch connectors. The core board has an FPGA (Artix-7 200T from Xilinx), onboard power modules, a double data rate synchronous dynamic random access memory, a quad serial peripheral interface flash, a 50 MHz crystal oscillator, and power management chips.

The base board uses 6 ADAS1128 chips, providing 768 input channels for reading Micromegas signals. For each channel, an Electro-Static Discharge (ESD) protection circuit is designed. The detector signals are integrated and converted to digital format within the ADAS1128 chip. The FPGA processes the data from the six ADAS1128s and transfers them via the communication interface as required. The communication interfaces of the baseboard comprise an optical transceiver and a USB 3.0 port. The optical fiber circuit is designed with a Clock and Data Recovery (CDR) chip, ADN2817, which can retrieve clock and serial data from the optical fiber link. The USB 3.0 circuit uses the CYUSB3014 chip designed by Cypress, facilitating single-board test.

 The number of fired detector strips for one cosmic-ray muon event typically ranges from about 2 to 10 in one direction. Based on test results, the maximum number of fired strips for a single muon event is 25. The data amount for a single muon event in one FEE board unit is approximately 150 bytes (50$\times$24 bits). 

\subsection{Introduction to DAQ Boards and Transmission Protocols}
\begin{figure}[t]
\centerline{\includegraphics[width=3.5in]{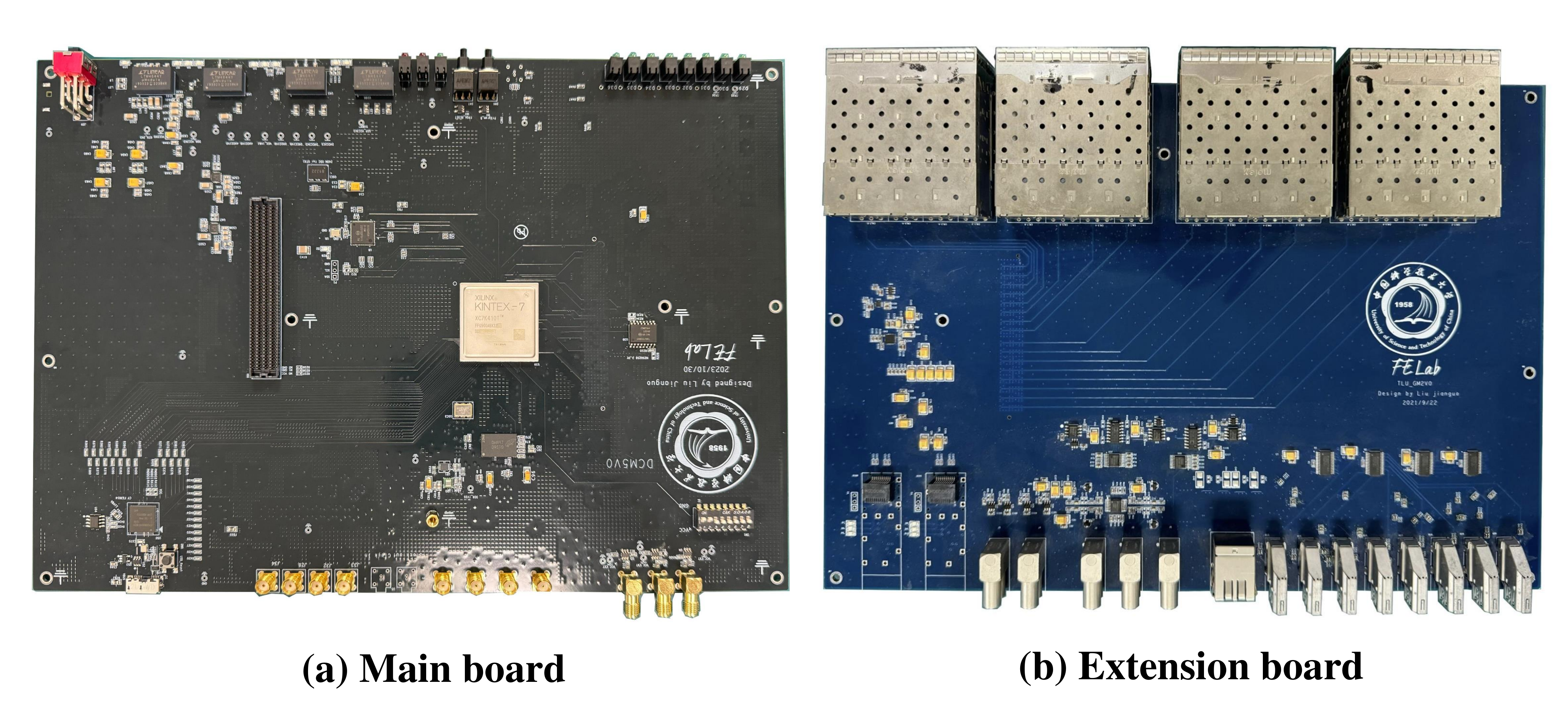}}
\caption{Photograph of the DAQ board.}
\label{fig5}
\end{figure}

The back-end electronics utilize the previously designed DAQ board shown in \figurename~\ref{fig5}. It consists of a main board and an extension board (following the terms used on \figurename~\ref{fig5}). The extension board is connected to the main board through the FPGA Mezzanine Card (FMC) interface. The main board uses an FPGA as the main control chip, which is mainly responsible for FEE configuration, data aggregation, clock distribution, and trigger elaboration. The extension board provides 32 optical fiber links for connecting FEE boards and offers various connectors for other external devices.

\begin{figure}[t]
\centerline{\includegraphics[width=3.5in]{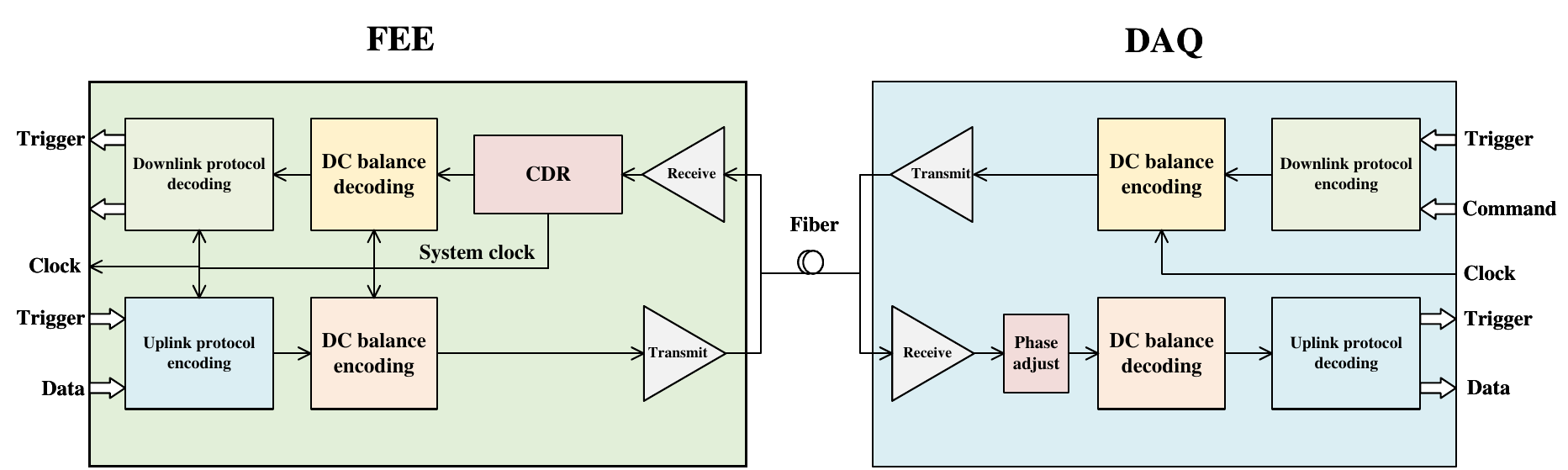}}
\caption{Diagram of communication between FEE and DAQ~\cite{bib:bib12}.}
\label{fig8}
\end{figure}

The DAQ board communicates with FEE boards via optical fibers using a custom protocol, as shown in \figurename~\ref{fig8}. The DAQ board sends the global reference clock of the system and configuration data with Direct Current (DC) balanced encoding to FEE, and FEE retrieves the original clock and data from the serial stream using a CDR chip. The original clock is processed by one of the internal Mixed-Mode Clock Managers (MMCMs) of the FPGA to produce the system clock. Data that have been time-multiplexed for transmission are split into trigger signals and configuration commands at the FEE end. 

The FEE generates “pre-trigger” signals and acquisition data, which are encoded and transmitted to the DAQ board over the same optical fiber. Since the clock source corresponding to the data stream is synchronized with the DAQ board, the DAQ board can catch the data with phase adjustment through FPGA logic.

\subsection{Trigger Modes of the System}
The electronics can operate in two trigger modes: self-trigger and external-trigger. In external-trigger mode, the DAQ board receives a trigger pulse from another system and sends it to the FEE boards. The FEE boards then transmit the data to the DAQ board.

\begin{figure}[t]
\centerline{\includegraphics[width=3.5in]{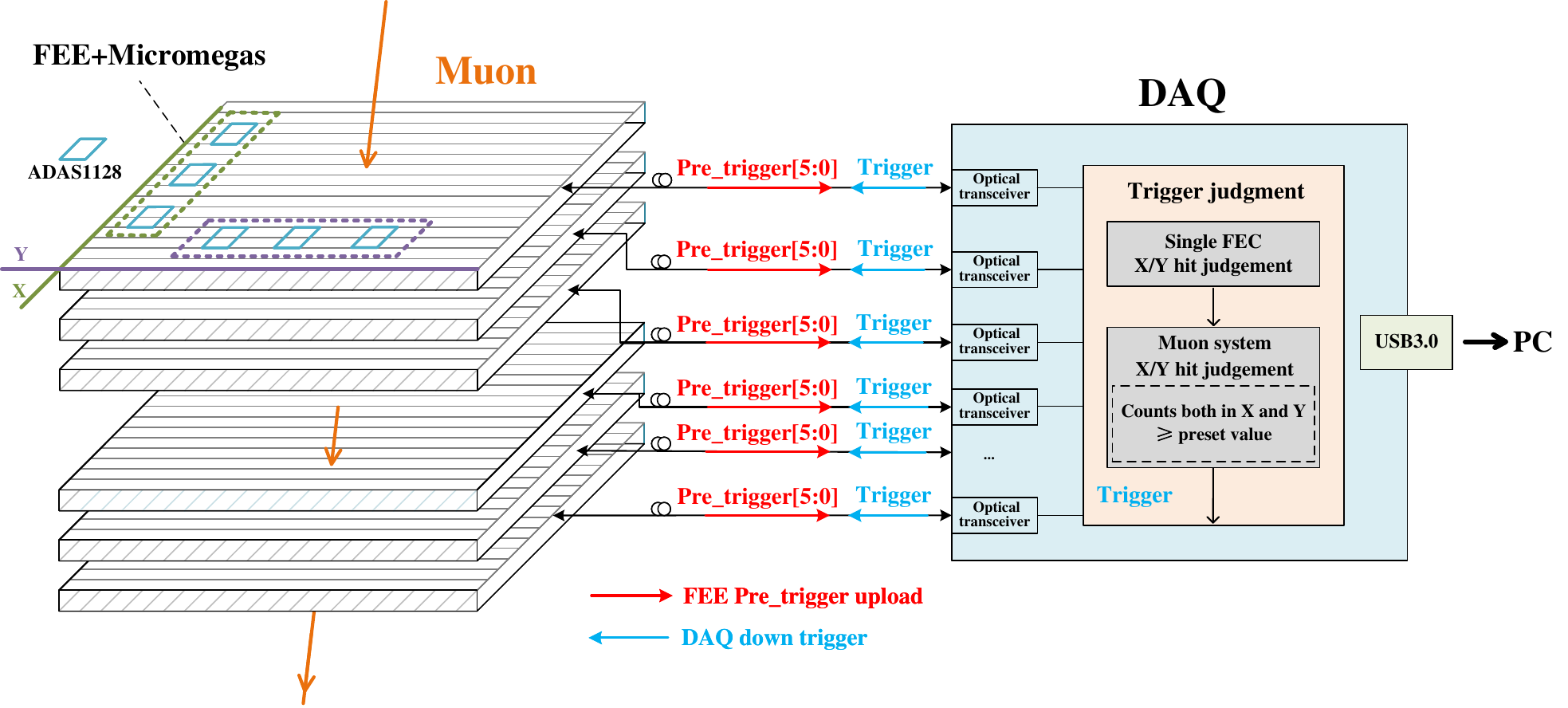}}
\caption{Schematic design of the self-trigger.}
\label{fig18}
\end{figure}

In self-trigger mode, the system will judge whether a valid muon is detected. The schematic design of the self-trigger is shown in \figurename~\ref{fig18}.
When a single chip detects that the signals from at least any two channels exceed a pre-defined threshold, the over-threshold flag signal of the chip is set to 1; otherwise, it remains at 0. 
After all the data of one acquisition period are judged, each FEE sends a "pre-trigger" word composed of 6 hit bits (one hit bit per ADAS1128 chip) to the DAQ board. 

The DAQ board combines the pre-trigger words of all FEE boards using FPGA logic to produce the valid trigger.
Firstly, it checks if there are hits in both the X and Y directions of the detector corresponding to each FEE. 
Then, it counts the number of hits in the X and Y directions of the whole system and tests if these exceed preset values. 
If the counts exceed the preset values, the DAQ board sends a trigger together with the trigger ID to all FEE boards. 

\subsection{Muon Imaging System Prototype}
\begin{figure}[t]
\centerline{\includegraphics[width=3.5in]{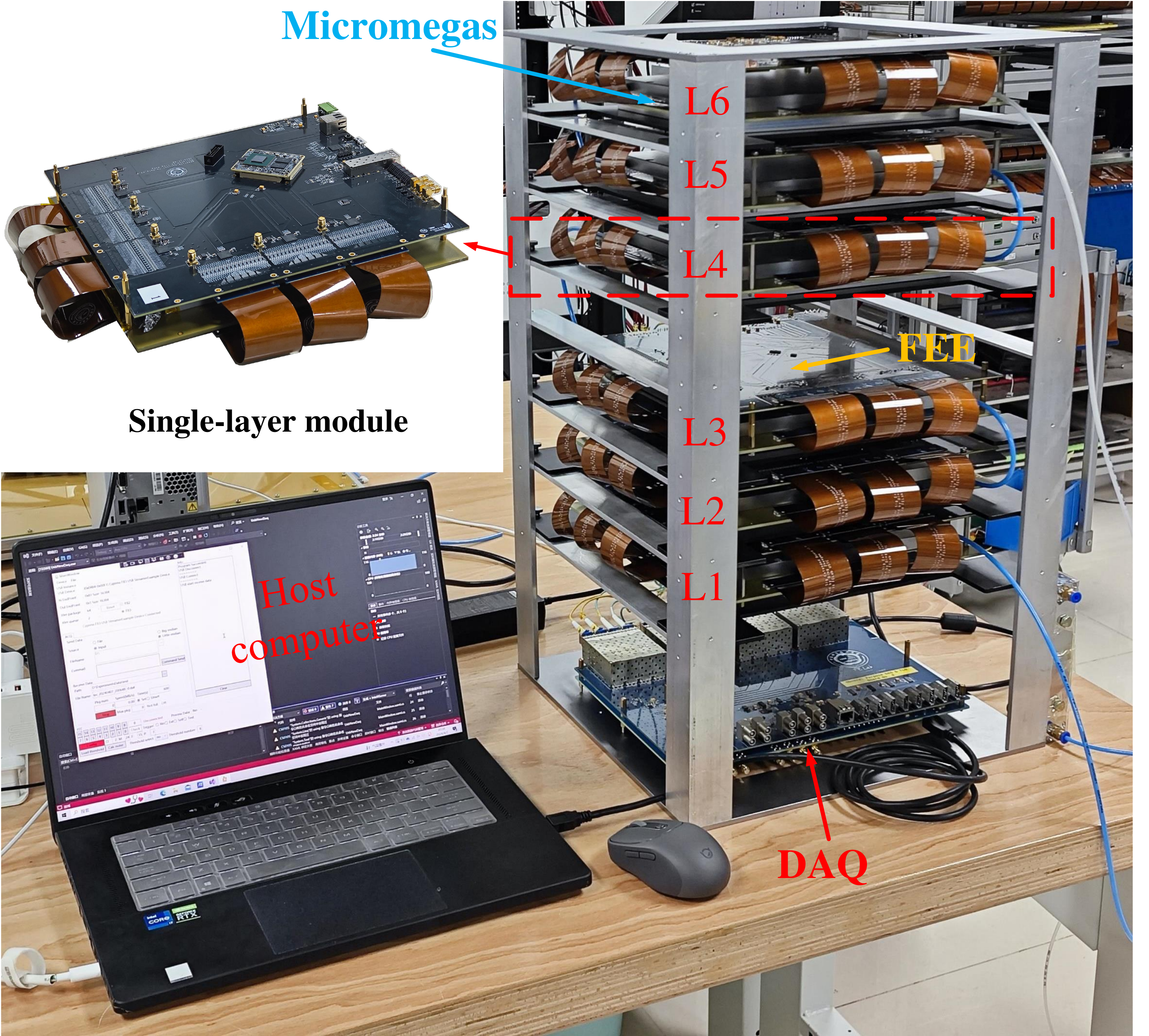}}
\caption{Photograph of muon imaging system prototype.}
\label{fig7}
\end{figure}

Our muon imaging system prototype consists of 6 layers of Micromegas detectors, 6 FEE boards, a DAQ board, a host computer, and high-voltage modules, as shown in \figurename~\ref{fig7}. Each FEE is connected to a 15 cm $\times$ 15 cm Micromegas detector through flexible printed circuit boards, and they are integrated in a single module. The entire system operates in self-trigger mode. The readout strips of each detector are grounded via 1 M$\Omega$ resistors, and the mesh voltage and drift electrode voltage are set to -550 V and -678 V, respectively. With this configuration, the gain of detectors is greater than 10$^4$\cite{bib:bib6}.

\section{Experimental Results}

\subsection{Performance of the Readout Electronics}
\begin{figure}[t]
\centerline{\includegraphics[width=3.5in]{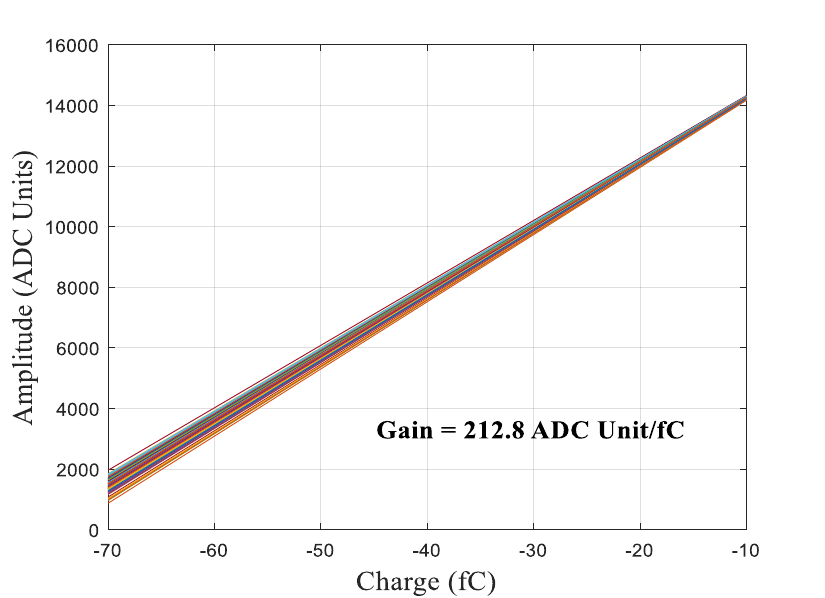}}
\caption{Channel response linearity of a single ADAS1128 chip.}
\label{fig9}
\end{figure}

In the channel response linearity test, the voltage signal generated by an external generator is converted to a charge signal by generating a step-down signal and sending it to a 1pF (5\% precision) capacitor in series. The gain and performance of 128 channels from one FEE are presented in \figurename~\ref{fig9}. The integral non-linearity is 2.8\%. 

\begin{figure}[t]
\centerline{\includegraphics[width=3.5in]{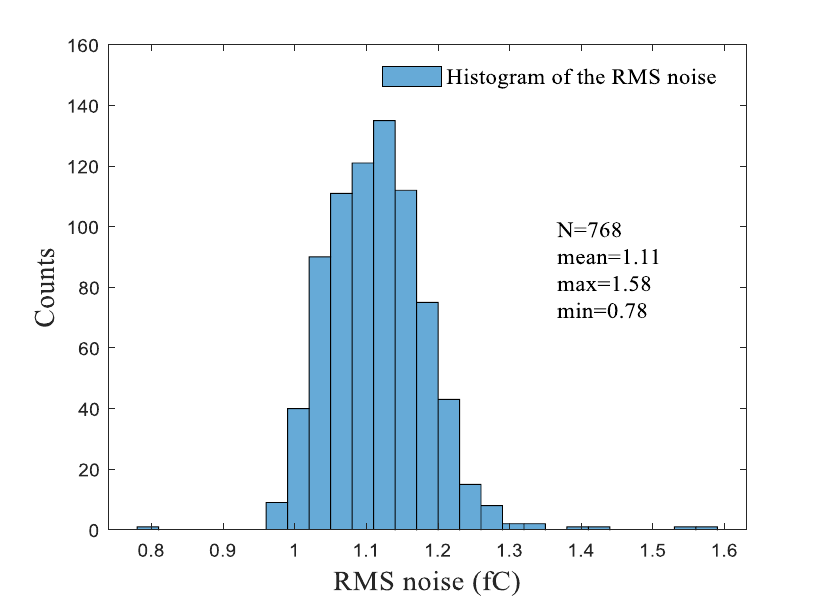}}
\caption{Distribution of the RMS noise of a single FEE connected to the detector.}
\label{fig10}
\end{figure}

The noise test is performed on FEE by recording the pedestal while the detector is connected. \figurename~\ref{fig10} shows the statistics of the equivalent input noise charge for all channels, with 95\% of the channels of the 6 ADAS1128 chips (768 channels) having a noise of less than 1.22 fC.

\subsection{Test with an X-Ray Source}
\begin{figure}[t]
\centerline{\includegraphics[width=3.5in]{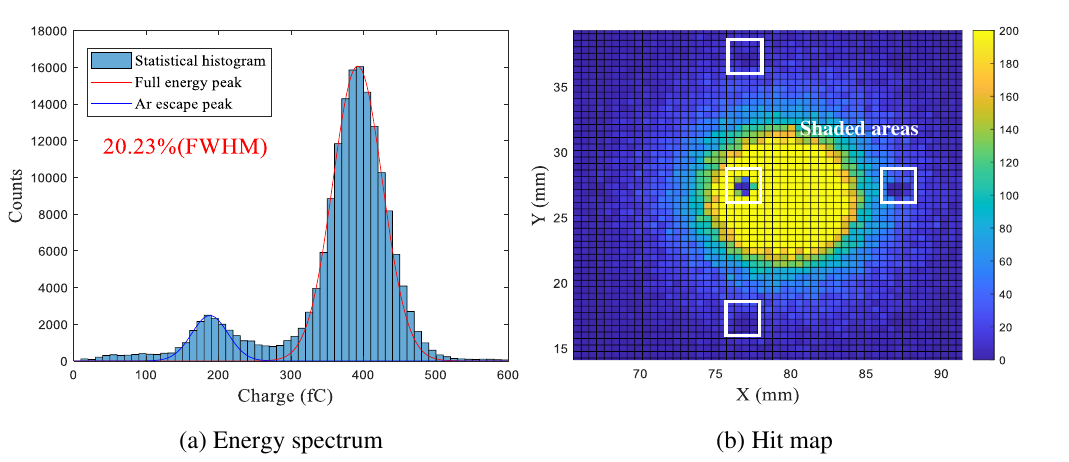}}
\caption{$^{55}\textnormal{Fe}$ radioactive test results.}
\label{fig11}
\end{figure}

The detector energy resolution tests are performed using a $^{55}\textnormal{Fe}$ radioactive source.
The energy spectrum and hit diagram are shown in \figurename~\ref{fig11}(a) and \figurename~\ref{fig11}(b) respectively. The peak position of the full energy peak is obtained by Gaussian fitting to be about 389.3 fC, and the energy resolution is 20.23\% FWHM (Full Width at Half Maximum) at 5.9 keV. The results of the energy resolution tests are consistent with the previous result~\cite{bib:bib9}.

There are shaded areas in the hit diagram, caused by the hot melt adhesive pads between the detector resistive electrode and the mesh electrode. These pads block the downward drifting electrons, preventing the electrons from drifting into the avalanche region. As a result, no inductive signal can be generated in this region.

\subsection{Cosmic Ray Muon Energy Deposition Result}
\begin{figure}[t]
\centerline{\includegraphics[width=3.5in]{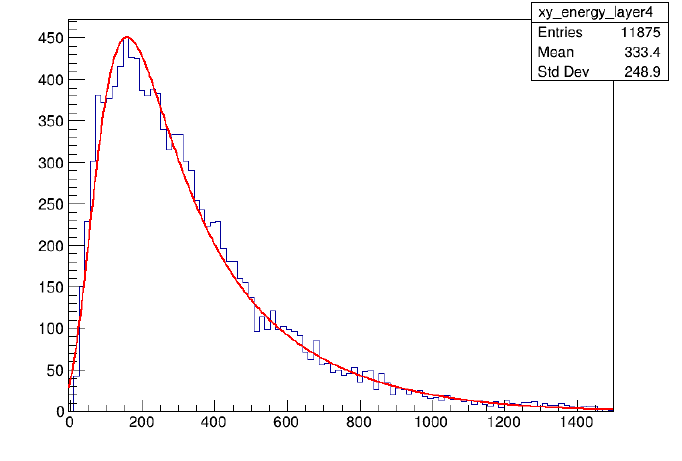}}
\caption{Measurements of the charge produced by cosmic ray muons in detectors.}
\label{fig15}
\end{figure}

Each cosmic ray muon will deposit energy in the drift area, generating induction signals on the readout strips in both the X and Y directions. The charge in each direction is obtained by summing up the charges on the detector strips corresponding to the reconstructed hit positions. Subsequently, the charges from the two directions are summed to obtain the cosmic ray energy deposition spectrum, as shown in \figurename~\ref{fig15}. From the energy spectrum, the charge distribution mainly follows the expected Landau distribution. After performing a Landau-Gaussian convolution fit using ROOT, the peak value is approximately 149.94 fC.

\subsection{Spatial Resolution Result}
\begin{figure}[t]
\centerline{\includegraphics[width=3.5in]{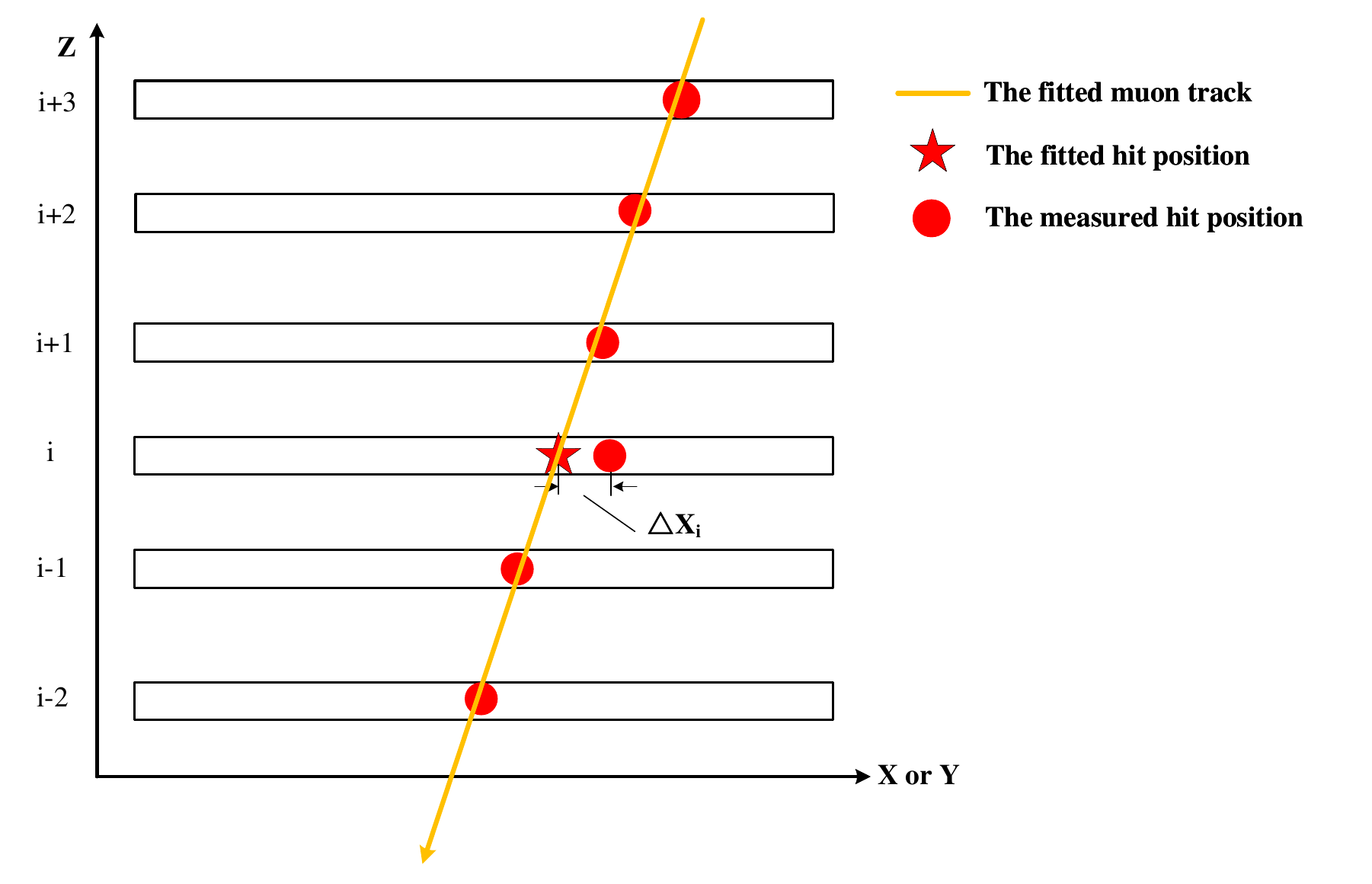}}
\caption{Schematic diagram of the method for calculating the spatial resolution.}
\label{fig16}
\end{figure}

 The method for calculating the spatial resolution of the muon track system is shown in \figurename~\ref{fig16}, with an example of calculating the spatial resolution of the i-layer. By using the hit positions of the other five layers as the reference track, the fitted hit position x$_{i_{fit}}$ can be obtained, where x$_{i_{fit}}$ is calculated by (1-3). Then, the offset $\Delta x_{i}$ was defined by (4).
By statistically analyzing a large number of muon tracks, the distribution of $\Delta x_{i}$ is obtained, and the standard deviation $\sigma$ ($\Delta x_{i}$) of the distribution is the spatial resolution of the layer detector.
\begin{equation}
x_{i_{fit}}=k_{x, i}\times z_{i}+b_{x, i}
\end{equation}
\begin{equation}
k_{x, i}=\frac{\sum_{j \neq i}^{n}\left(z_{j}-\bar{z}\right) x_{j_{h i t}}}{\sum_{j \neq i}^{n}\left(z_{j}-\bar{z}\right)^{2}}
\end{equation}
\begin{equation}
b_{x,i}=\bar{x}-k_{i}
\end{equation}
\begin{equation}
\Delta x_{i}=x_{i_ {f i t}}-x_{i_{ h i t}}
\end{equation}

\begin{figure}[t]
\centerline{\includegraphics[width=3.5in]{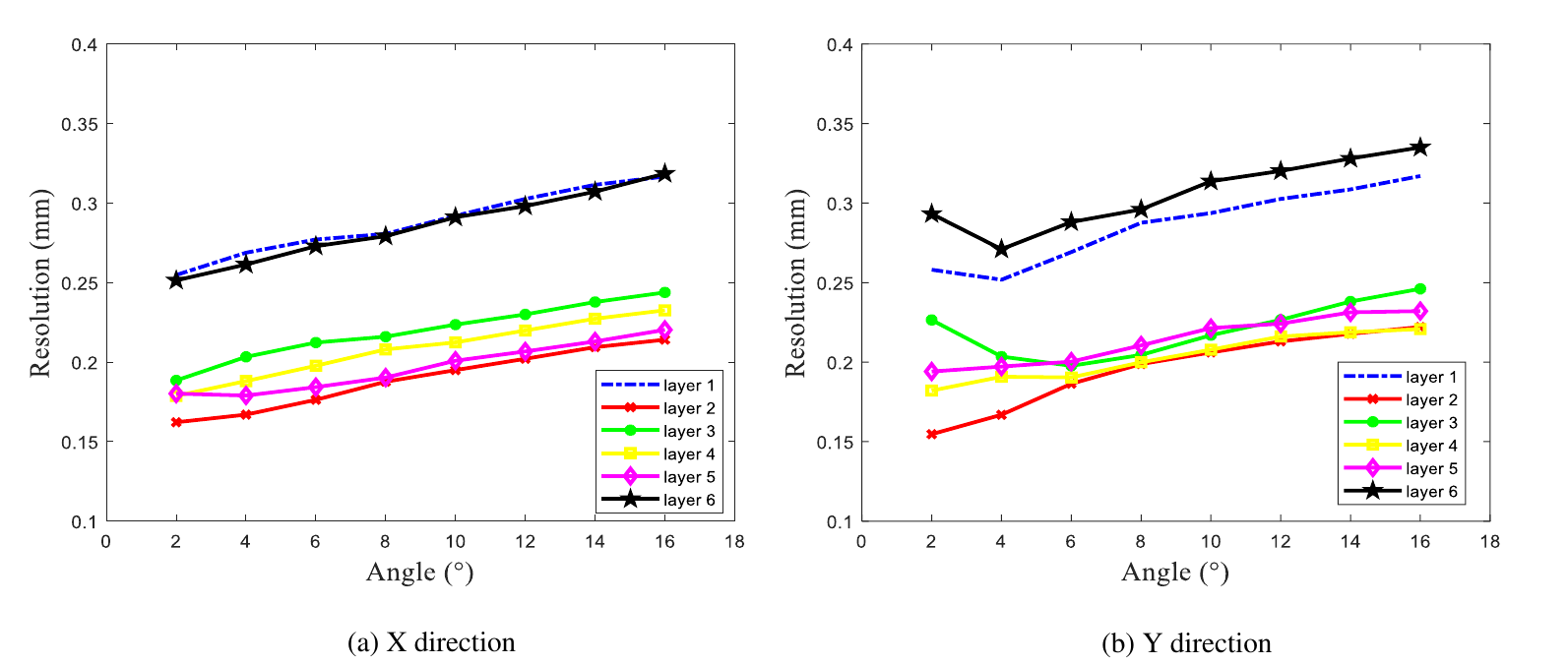}}
\caption{Spatial resolution of Micromegas detectors.}
\label{fig17}
\end{figure}

The spatial resolution of the Micromegas detector in the X and Y directions is shown in \figurename~\ref{fig17}. When calculating instances from different angles separately, the uncertainty in the initial ionization position increases as the incident angle of the muon increases. For 0 $ \sim $ 5$^{\circ}$ incident muons, the spatial resolution of the detector is better than 200 $\mu$m. In the detector anode PCB (Printed Circuit Board), X direction readout strips are located above the Y direction, and X direction readout strips can collect more charge. Therefore, the spatial resolution in the X direction is slightly better than in the Y direction.

\subsection{Efficiency Result}
The detection efficiency is defined as the ratio of effective events recorded by the target detector (i-layer) to coincidence events of the reference tracker (the remaining five layers of detectors), where the effective events of the target detector are selected with the condition that $\Delta x_{i} < 10 \times \sigma$ ($\Delta x_{i}) $. 
The detection efficiency in the X and Y directions of detectors is shown in Table \ref{tab:wang.t1}. Results show that the efficiency of most of the 6-layer Micromegas detectors is over 95\%.

\begin{table}[htbp]
\centering
\caption{Efficiency of Micromegas Detectors}
\label{tab:wang.t1}
\begin{tabular}{ccccccc} 
\toprule 
 & Layer-1 & Layer-2  & Layer-3 & Layer-4 & Layer-5 & Layer-6 \\ 
\midrule 
X & 98.15\% & 97.75\% & 98.49\% & 98.01\% & 97.99\% & 97.50\% \\ 
Y & 94.62\% & 95.96\% & 98.25\% & 97.71\% & 97.29\% & 97.01\% \\ 
\bottomrule 
\end{tabular}
\end{table}

\subsection{Muon Scattering Imaging Result}
\begin{figure}[t]
\centerline{\includegraphics[width=3.5in]{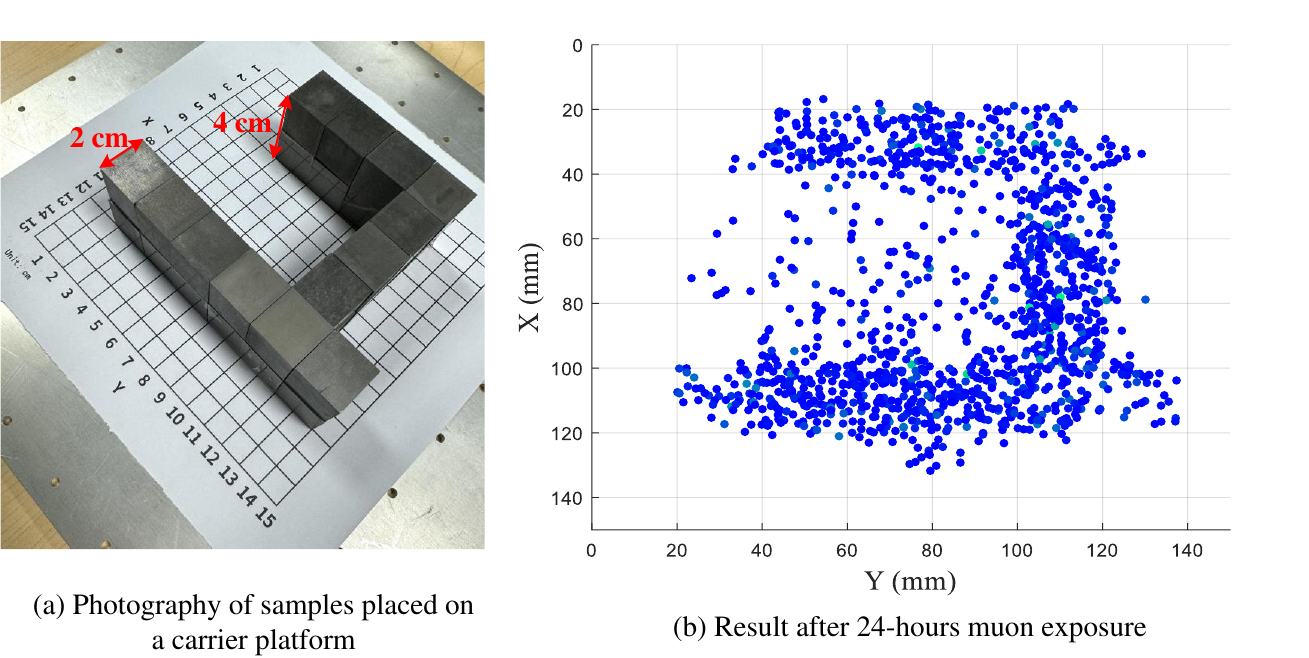}}
\caption{Muon scattering imaging results.}
\label{fig19}
\end{figure}

\figurename~\ref{fig19} shows the results of imaging blocks of tungsten (2 cm wide and 4 cm thick) placed on a carrier platform to form the letter $\mu$. After accumulating 24 hours of data, the imaging result was reconstructed using the Point of Closest Approach (PoCA) algorithm incorporating the k-Nearest Neighbors (k-NN) algorithm\cite{bib:bib14}. The reconstructed image shown in \figurename~\ref{fig19}(b) shows that our muon imaging system prototype can reconstruct the boundaries of the objects used in this test.

\section{Conclusion and Discussion} 
Under the scenario of low-rate cosmic ray muon imaging applications, a highly integrated FEE for reading out Micromegas detectors based on the current readout chip ADAS1128 was designed in this work. The performance of the readout electronics can satisfy the requirements of Micromegas detectors used in muon imaging. The RMS noise of the readout electronics is about 1.22 fC. The spatial resolution of the detectors is about 200 $\mu$m for small-angle incident muons. In addition, the constructed muon imaging system prototype was used for scattering imaging. The system prototype can reconstruct the boundaries of sufficiently massive objects (e.g., 4 cm thick tungsten blocks) with a width of 2 cm.

The system prototype presented in this paper indicates that reading out Micromegas using a commercial current readout chip is a promising solution for muon imaging.
For applications with low occupancy rates, the ADAS1128 chip can meet the readout requirements. The high level of integration of this device and its small size are advantageous for designing large-scale muon imaging readout systems. For large-area Micromegas detectors, longer readout strips increase the capacitance of the detector, which may lead to higher noise measured by FEE. We designed a new FEE following the same electronic structure and tested the readout performance of the 40 $\times$ 40 cm$^2$ Micromegas detector. The test results show that the performance of 40 $\times$ 40 cm$^2$ detectors is consistent with that of 15 $\times$ 15 cm$^2$ detectors. Therefore, the ADAS1128 chip can read out larger-area Micromegas detectors.

\end{document}